\documentclass{sig-alternate}
\pdfoutput=1
\usepackage[utf8]{inputenc}
\usepackage{graphicx}
\usepackage{caption}
\usepackage{color}
\usepackage{hyperref}
\usepackage{paralist}
\usepackage{balance}
\usepackage{enumitem}
\usepackage{microtype}
\usepackage{bm}
\usepackage{subfig}

\makeatletter
\def\@copyrightspace{\relax}
\makeatother

\newcommand{\paragraphcustom}[1]{\noindent\textbf{#1.}}
\newcommand{\savespace}[0]{\vspace{-1.25em}}
\definecolor{darkgreen}{RGB}{0, 128, 0}

\begin{document}
\title{Improving Reachability \\and Navigability in Recommender Systems}

\numberofauthors{3}
\author{
\alignauthor
Daniel Lamprecht\\
       \affaddr{KTI, Graz University of Technology}\\
       \affaddr{Graz, Austria}\\
       \email{daniel.lamprecht@tugraz.at}
\alignauthor
Markus Strohmaier\\
       \affaddr{GESIS and University of Koblenz-Landau}\\
       \affaddr{Cologne, Germany}\\
       \email{strohmaier@uni-koblenz.de}
\alignauthor
Denis Helic\\
       \affaddr{KTI, Graz University of Technology}\\
       \affaddr{Graz, Austria}\\
       \email{dhelic@tugraz.at}
}
\maketitle

\begin{abstract}
In this paper, we investigate recommender systems from a network perspective and investigate recommendation networks, where nodes are items (e.g., movies) and edges are constructed from top-N recommendations (e.g., related movies). In particular, we focus on evaluating the reachability and navigability of recommendation networks and investigate the following questions: (i) How well do recommendation networks support navigation and exploratory search? (ii) What is the influence of parameters, in particular different recommendation algorithms and the number of recommendations shown, on reachability and navigability? and (iii) How can reachability and navigability be improved in these networks? We tackle these questions by first evaluating the reachability of recommendation networks by investigating their structural properties. Second, we evaluate navigability by simulating three different models of information seeking scenarios. We find that with standard algorithms, recommender systems are not well suited to navigation and exploration and propose methods to modify recommendations to improve this. Our work extends from one-click-based evaluations of recommender systems towards multi-click analysis (i.e., sequences of dependent clicks) and presents a general, comprehensive approach to evaluating navigability of arbitrary recommendation networks. 
\end{abstract}


\section{Introduction}
An important use case of a recommender system is its ability to support browsing and navigation behavior. For example, we know that users enjoy perusing item collections without the immediate intention of making a purchase \cite{herlocker2004evaluating}. Flickr users have been found to predominately discover new images via \emph{social browsing} \cite{lerman2007social}. For platforms where users immediately consume content, such as YouTube, recommendations serve the use case of \emph{unarticulated want}, and are therefore a crucial part of user experience \cite{davidson2010youtube}. More generally, some users prefer navigation to direct search even when they know the target \cite{teevan2004perfect}. In such exploratory scenarios, the knowledge gained along the way provides context and aids in learning and decision-making \cite{resnick1997recommender, marchionini2006exploratory}.

\begin{figure}[t!]
    \centering
    \includegraphics[width=\linewidth]{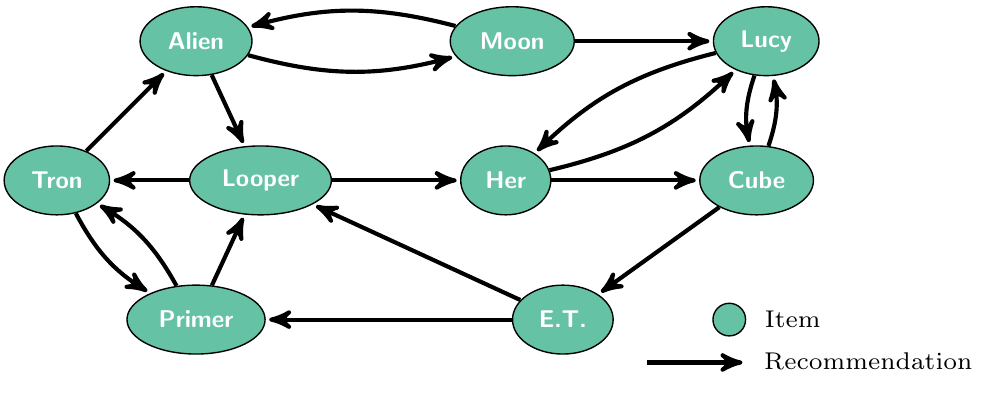}
    \caption{\textbf{Recommendation Network.} Recommender systems implicitly form recommendation networks, where nodes are items (e.g., movies) and edges are directed hyperlinks between related items. The illustration shows a scenario where two recommendations are available for each movie.}
\label{fig:recnet}
\savespace
\end{figure}

Historically, there is only little research to assess the ability of recommender systems to aid such navigation. Yet, hyperlinks in a recommender system are, by their very conception, meant to be navigated and used for search and exploration tasks. While a few studies have already looked at recommendation networks (see Figure \ref{fig:recnet} for an example) and provided first important insights into the nature and structure of these networks \cite{cano_topology_2005, celma_new_2008, seyerlehner_browsing_2009}, there is no systematic approach to evaluating navigability both statically (reachability) and dynamically (navigability).\\

\paragraphcustom{Research questions} We address the problem of evaluating both reachability and navigability of recommendation networks by investigating three research questions:
\begin{enumerate}
 \item How well do recommendation networks support navigation and exploratory search?
\item What is the influence of parameters, in particular different recommendation algorithms and the number of recommendations shown, on reachability and  navigability?
\item How can reachability and navigability in recommender systems be improved?
\end{enumerate}

\paragraphcustom{Approach} We study two types of recommendation networks: \begin{inparaenum}[(i)]
 \item networks where links are generated from \emph{collaborative filtering} algorithms (using explicit user ratings), and
\item networks where links are generated from \emph{content} (using text similarity).
\end{inparaenum}
We then use a two-level approach to evaluating reachability and navigability:
\begin{enumerate}
\item \textbf{Reachability}: We evaluate reachability of recommendation networks by looking at the network topology: components, clustering, path lengths and bow-tie structure. Analyzing the topological nature of recommendation networks provides us with first insights into the extent to which items are \emph{reachable}.
\item \textbf{Navigability}: We then evaluate practical navigability of recommendation networks using three different navigation models established in the literature:
\begin{inparaenum}[(i)]
 \item \emph{Point-to-Point Search} \cite{kleinberg2000} as an example of goal-oriented navigation with a single fixed goal,
\item \emph{Berrypicking} \cite{bates1989design} as an example of goal-oriented navigation with multiple and variable goals, and
\item \emph{Information Foraging} \cite{pirolli2007} as an example of exploration.
\end{inparaenum}\\
\end{enumerate}

\paragraphcustom{Contributions} Our contributions are three-fold:

First, we present a general approach for evaluating navigability of arbitrary recommendation networks via both topological analysis and navigation models (simulation), and demonstrate the feasibility of this approach by applying it to actual recommendation networks.

Second, we find that the recommender systems we investigate are poorly connected and not very well-suited for navigation and exploration. On our datasets, we find that recommendation networks generated by collaborative-filtering algorithms perform better for most navigation scenarios. While this suggests that collaborative filtering is a better choice for the creation of navigable recommendations on our datasets, we leave the task of applying our approach to other (real-world) recommendation networks and datasets to future work.

Third, we propose a series of simple changes to recommendation algorithms to introduce \emph{navigational diversity} and demonstrate how they help to overcome the reachability problem in recommender networks.

\section{Related Work}
\label{sec:related}
\paragraphcustom{Network Analysis} Ever since Milgram's small world experiments \cite{milgram67}, researchers have been making efforts to understand \emph{navigability} and in particular \emph{efficient navigation} in networks. Kleinberg \cite{kleinberg2000a,kleinberg2000} and Watts \cite{watts02} formalized the property that a navigable network requires short paths between all (or almost all) nodes \cite{kleinberg2001}. Formally, such a network has a low diameter bounded by a polynomial in $log(n)$, where $n$ is the number of nodes in the network, and a giant component containing almost all the nodes exists~\cite{kleinberg2001}. In other words, because the majority of network nodes are connected, it is possible to reach all or almost all of the nodes, given global knowledge of the network. This property is referred to as \emph{reachability}. The low diameter and the existence of a giant component constitute necessary topological conditions for network navigability. In this paper we apply a set of standard network-theoretic measures including distribution of component sizes and component structure (via the bow tie model) to assess if a network satisfies them.

Kleinberg also found that an \emph{efficiently navigable} network possesses certain structural properties that make it possible to design efficient local search algorithms (i.e., algorithms that only have local knowledge of the network) \cite{kleinberg2001,kleinberg2000}. The delivery time (the expected number of steps to reach an arbitrary target node) of such algorithms is then sub-linear in $n$. In this paper, we investigate the efficient navigability of networks through the simulation of a range of search and navigation models.

\paragraphcustom{Recommender Systems}
Initially, recommender systems were mostly evaluated in terms of recommendation accuracy. More recently, the importance of evaluation metrics beyond accuracy has been identified \cite{herlocker2004evaluating, shani2011evaluating}. Approaches such as diversity (e.g., \cite{boim2011diversification}), novelty (e.g., \cite{nakatsuji2010classical}), and serendipity, which are thought to be orthogonal to the traditional accuracy-based evaluation measures, have been found to increase user satisfaction \cite{bookcrossing}. Recommender systems have been found to show a filter bubble effect (even though following recommendations actually \emph{lessened} the effect) \cite{nguyen2014exploring}. Diversification of recommendations can be an effective means of increasing the spectrum of recommendations users are exposed to.

In terms of reachability, the static topology of recommendation networks has been studied for the case of music recommenders. Their corresponding recommendation networks have been found to exhibit heavy-tail degree distributions and small-world properties \cite{cano_topology_2005}, implying that they are efficiently navigable with local search algorithms. 
Seyerlehner et al. studied sources (nodes that are never recommended) in music recommendation networks \cite{seyerlehner_limitations_2009} and found that the fraction of sources remained constant independent of the recommendation approach and the network size. This indicates that recommendation networks generally suffer from reachability problems.
Celma and Herrera \cite{celma_new_2008} found that collaborative filtering on \emph{last.fm} led to recommendation networks that are prone to a popularity bias, with recommendations biased towards popular songs or artists. They also found that collaborative filtering provided the most accurate recommendations, while at the same time this made it harder for users to navigate to items in the long tail. A hybrid approach and content-based methods provided better novel recommendations. These results suggest that a trade-off exists between accuracy and other evaluation metrics. Mirza et al. \cite{mirza2003studying} proposed to measure reachability in the bipartite recommendation graph of users and items as an evaluation measure.

A simple method to improve reachability is to select recommendations specifically for their target in the network, which has been proposed by Seyerlehner et al. \cite{seyerlehner_browsing_2009}. In this work, we improve on this method by proposing a method that not only improves reachability but also ensures the relevance of the selected recommendations.

While these analyses have shown certain topological properties such as heavy-tail degree distributions and small-world properties \cite{cano_topology_2005}, we know very little about the dynamics of actually using recommendations to find navigational paths through a recommender system.

\section{Methods \&  Experimental Setup}
\label{sec:methods}

In the following, we briefly sketch our approach for assessing network navigability and describe the datasets and the methods we applied to generate recommendation networks. All datasets
are publicly available.

\subsection{Item Datasets}
We look at two examplary types of items for our experiments: movies from MovieLens and books from BookCrossing.

\noindent\textbf{MovieLens} is a film recommender system by the University of Minnesota. For this work, we used the MovieLens dataset consisting of one million ratings \footnote{\href{http://grouplens.org/datasets/movielens/}{http://grouplens.org/datasets/movielens}} from $6,000$ users on $4,000$ movies, where each user had rated at least 20 movies. From this, we used the $3,640$ movies that had corresponding Wikipedia articles.

\noindent\textbf{BookCrossing} is a book exchange platform. For this work, we used a 2005 crawl of the website \cite{bookcrossing}. As a preprocessing step, we filtered out implicit ratings and users which had rated fewer than 20 books, leaving us with $110,610$ books. From these, we randomly sampled books until we were able to match $3,640$ books to their corresponding Wikipedia articles. This left us with two datasets of the same size.

\paragraphcustom{Mapping to Wikipedia articles} For the computation of content-based recommendations, we mapped the movie and book titles to corresponding articles in the English Wikipedia and extracted the textual content. For the mapping we made use of the naming conventions for movies and books used on Wikipedia. For movies, the article title is generally the movie title itself (e.g., \emph{Alice in Wonderland}). However, in case the title also refers to other media (such as the Lewis Caroll book), the Wikipedia title will be \emph{Alice in Wonderland (film)}. Should several movies with the same title exist, the Wikipedia title will be \emph{Alice in Wonderland (1951 film)}. The same conventions hold true for book titles. 

\captionsetup[subfigure]{labelformat=empty}
\newlength{\figheight}
\setlength{\figheight}{2.65cm}
\begin{figure*}[t!]
  \centering
  \makebox[80pt]{\raisebox{35pt}{(a) MovieLens}}%
  \subfloat[]{\includegraphics[height=\figheight]{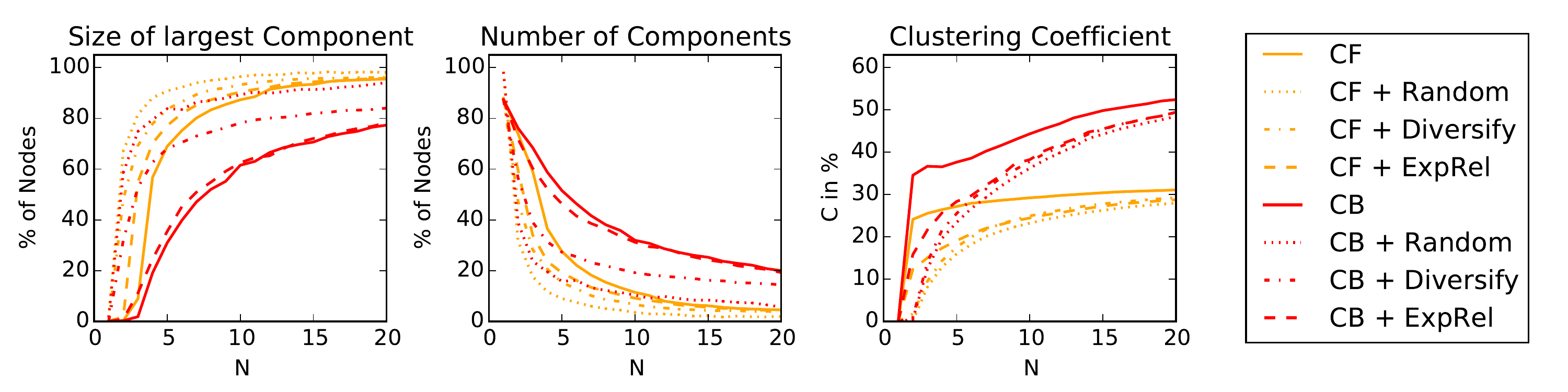}\includegraphics[height=\figheight]{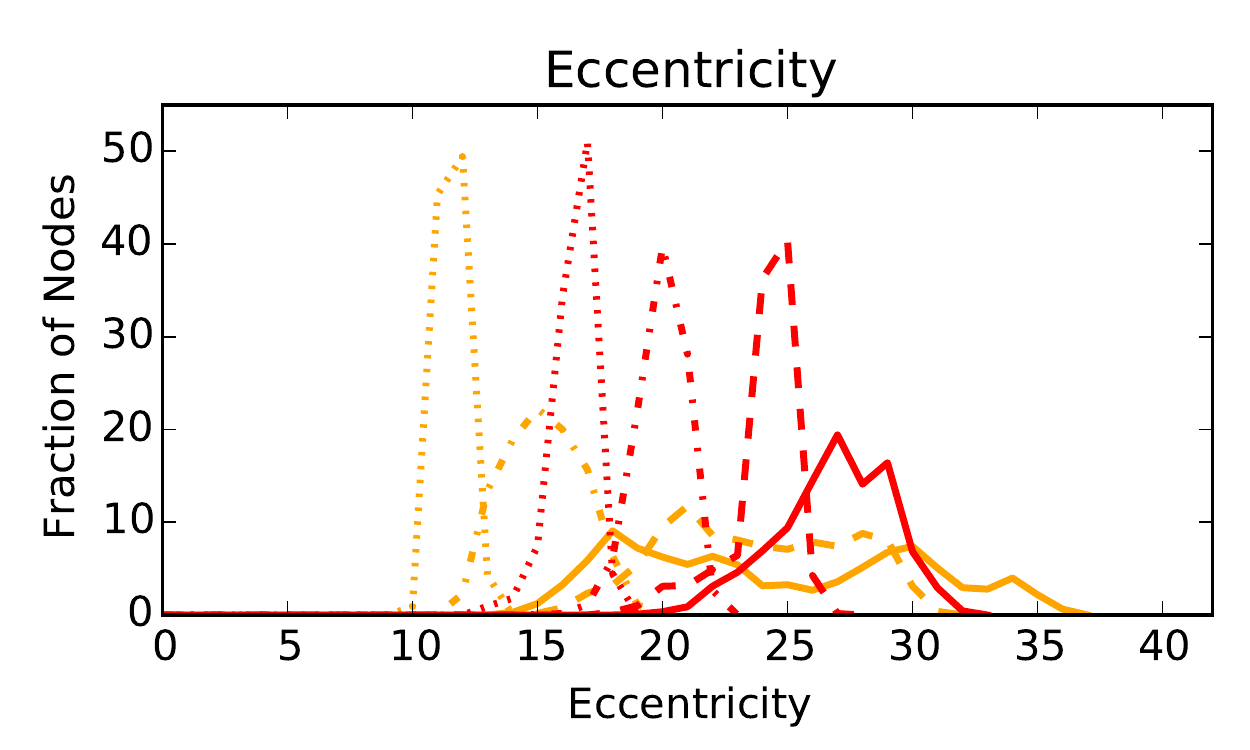}}
  \\[-2.5em] 
  \makebox[80pt]{\raisebox{35pt}{(b) BookCrossing}}%
  \subfloat[]{\includegraphics[height=\figheight]{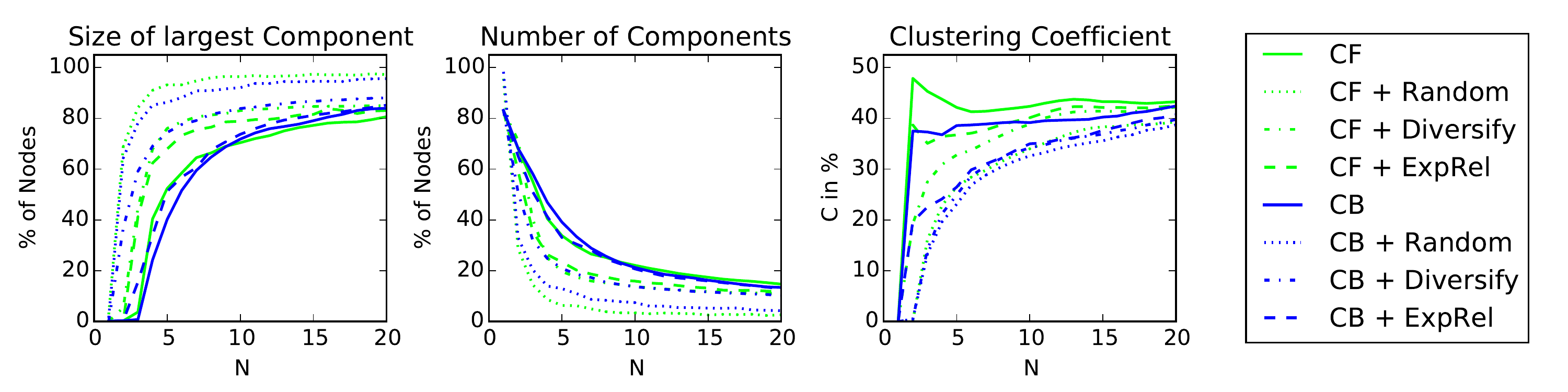}\includegraphics[height=\figheight]{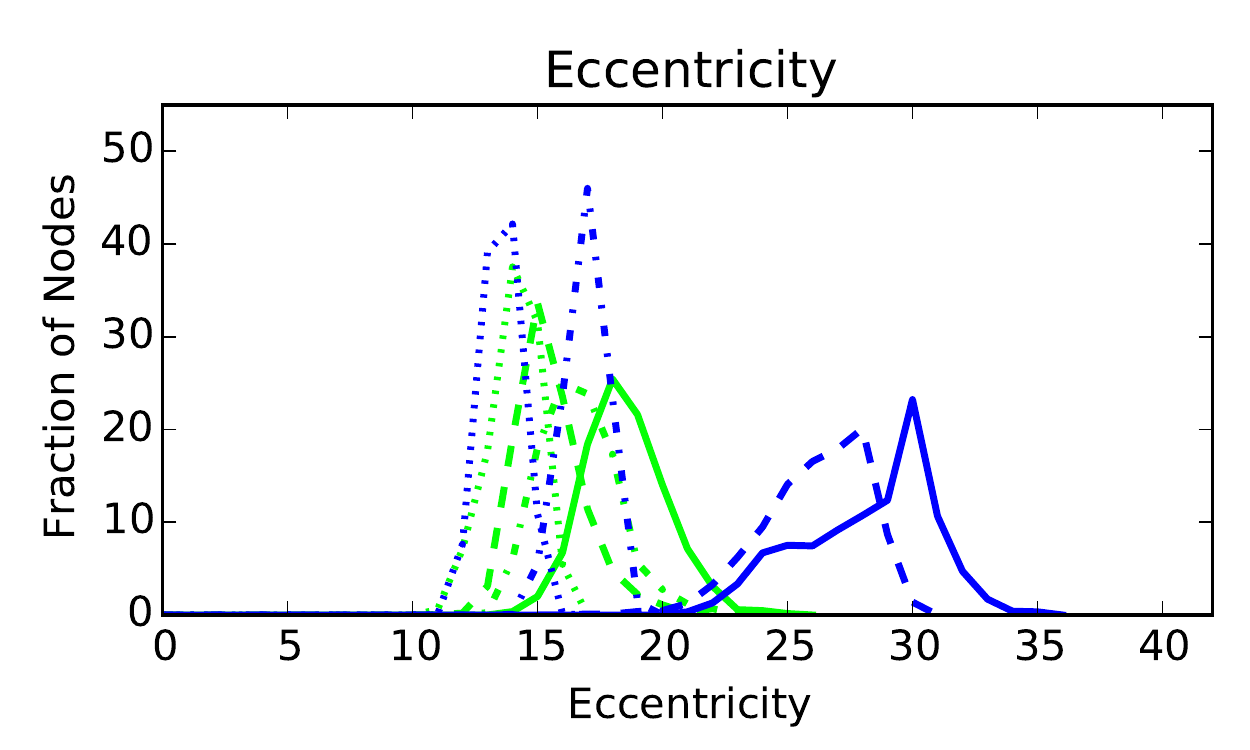}}
  \vspace{-2em}
  \caption{\textbf{Reachability analysis in terms of components, clustering and eccentricity.} The first two columns show the size of the largest strongly connected component in the graphs and the number of components present. We find that recommendation networks are not well-connected, but this improves with the application of diversification. The third column shows the clustering coefficient. Diversification led to larger components and lower clustering and thus potentially has the downside of potentially making navigability harder. The fourth column depicts the distribution of eccentricities (the maximum shortest distances) between all pairs of nodes in the largest strongly connected component for $N=5$ recommendations. While eccentricity values are shorter for collaborative filtering (CF) than for content-based (CB) recommendation networks, distances are too long for navigation in real-world systems in both types of networks. Diversification approaches help in reducing path lengths.}
  \label{fig:topology}
  \savespace
\end{figure*}

\paragraphcustom{Genres} We clustered items based on genre information, which for MovieLens was supplied with the dataset. For BookCrossing, we used information provided by Google Search. As of early 2015, querying Google Search for e.g., \emph{Alice in Wonderland genres} produces an infobox containing a range of standardized genre affiliations, which we extracted and added our dataset. A manual inspection of 50 randomly sampled books revealed that this was highly accurate in terms of precision.
\subsection{Building Recommendation Networks}

We calculated non-personalized collaborative-filtering and content-based recommendations for both datasets in the following way: For the given set of items $I$ and a similarity measure, we compute the pairwise similarities for all pairs of items $i$ and $j$. For each item $i \in I$, we define the set of the top-N most similar items to $i$ as $L_{i, N}$. We then create a top-N recommendation network $G\left(V, N, E\right)$, where $V = I$ and $E = \{ \left(u, v\right) | u \in I, v \in L_{u, N}\}$. We investigated values for $N$ in $\left[1, 20\right]$, which we consider a plausible range for recommender systems. This method leads to recommendation networks with constant outdegree and varying indegree--representing a typical setting for top-N recommendation networks.

\paragraphcustom{Collaborative Filtering Recommendations (CF)} We used the user-rating matrices associated with the datasets to calculate non-personalized collaborative filtering recommendations. We considered the ratings for each item as a vector, from which we then calculated cosine similarities to all the other item vectors and presented the top-N most similar vectors as recommendations.

\paragraphcustom{Content-based Recommendations (CB)} As a second recommendation approach, we calculated simple content-based recommendations. For this purpose, we computed the TF-IDF features on the Wikipedia articles corresponding to the items and then calculated the (non-personalized) cosine similarities between these feature vectors and used the top-N recommendations.

\subsection{Improving Reachability and Navigability through Diversification}

We experimented with three simple approaches to improve reachability and navigability. We applied these approaches to all top-N recommendation sets $L_{i, N}$ for $N \geq 2$ and used them to find a replacement for the least similar recommendation in $L_{i, N}$.

\begin{itemize}[leftmargin=*]
  \item \textbf{Random.} Random graphs generally exhibit a small diameter. By replacing the last recommendation with a random item, this algorithm served as an upper bound in terms of achievable reachability improvement when replacing a single recommendation for every item.
  \item \textbf{Diversify.} This approach aimed at diversifying the recommendation list. The replacing recommendation was chosen among the overall top-50 recommendations for the given item as the one maximizing the pairwise dissimilarity to the items already in the list--a procedure referred to as \emph{diversify} by Ziegler et al. \cite{bookcrossing}. The similarity measures in this case were the cosine similarity of the rating vectors or TF-IDF vectors.
  \item \textbf{Expanded Relevance (ExpRel)} This method chose the additional recommendation among the overall top-50 recommendations for the given item as the one maximizing the one-step expanded relevance, as proposed by Küçüktunç et al. \cite{kuccuktuncc2013diversified}. The algorithm ranks nodes by taking into account both the relevance of the potential node as well as the fraction of its one-hop neighborhood that is not already directly reachable via other recommendations. This approach aims at adding diversity by providing a connection into a useful area of the recommendation graph.
\end{itemize}

\section{Reachability}
\label{sec:topology}
In the following, we present the topological characteristics of the recommendation networks introduced in the previous section. We will focus on analyzing
\begin{inparaenum}[(i)]
  \item effective reachability in terms of components and local clustering,
  \item efficient reachability in terms of path lengths (eccentricity) and
  \item reachable partitions of the graph in terms of a bow tie model analysis. \end{inparaenum}
This provides us with insights into the topological nature of these networks and with general clues for the high-level navigation characteristics of such structures.

\subsection{Effective Reachability}
\paragraphcustom{Description} The analysis of the size of the largest connected component and the distribution of component sizes and gives a direct answer to the \emph{reachability} of a recommendation network. Determining the size of the largest strongly connected component is related to \emph{catalog coverage}, but goes beyond this in that it measures the size of the largest subset of items that are not only recommended but also mutually reachable. The local clustering coefficient provides insight into the local topology and can give us hints about how globally observed components emerge. It is computed as
\begin{equation}
C = \frac{1}{|V|}\sum_{i \in V} \frac{|\{(j, k) \in E | j,k \in \Gamma(i)\}|}{|\Gamma(i)|\, (|\Gamma(i)| - 1)},
\end{equation}
where $\Gamma(i)$ is the set of nodes reachable from $i$. For example, strong local clustering with a high number of small components and the absence of a giant component would indicate that the network consisted of a high number of isolated \emph{caves} that are not connected to each other \cite{Watts99}. In terms of recommendation systems these can be groups of items that mutually recommend each other. This analysis can provide us with explanations for observed phenomena.

\paragraphcustom{Results and Interpretation} The size of the largest strongly connected components in the networks grew with $N$ (the number of recommendations pointing away from items)--see Figure \ref{fig:topology}. For $N > 7$, the largest component contained at least 50\% of the nodes for all networks. Collaborative filtering led to larger components than the content-based approach, with MovieLens having the largest component thereof.

In real-world examples, the number of immediately visible recommendations typically lies between four and twelve. For instance, Amazon recommends between five and eight items (depending on screen resolution), YouTube recommends twelve videos and IMDb lists six related films. If our examples generalize to these datasets, this comparison shows that standard recommendation approaches allow users to explore only around half the network by browsing. Even for 20 recommendations, a great number of components still exists in the network in the range of 5--20\% of the network nodes (e.g., for our networks with 3,640 nodes, 10\% components translate to 364 components).

\begin{figure*}[t!]
    \centering
  \makebox[11pt]{\raisebox{30pt}{(a)}}%
  \subfloat[]{\includegraphics[width=\textwidth]{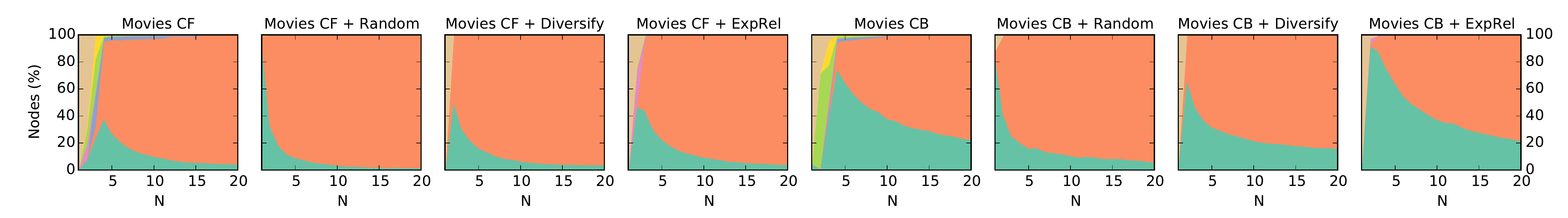}}
  \\[-3em] 
  \makebox[11pt]{\raisebox{30pt}{(b)}}%
  \subfloat[]{\includegraphics[width=\textwidth]{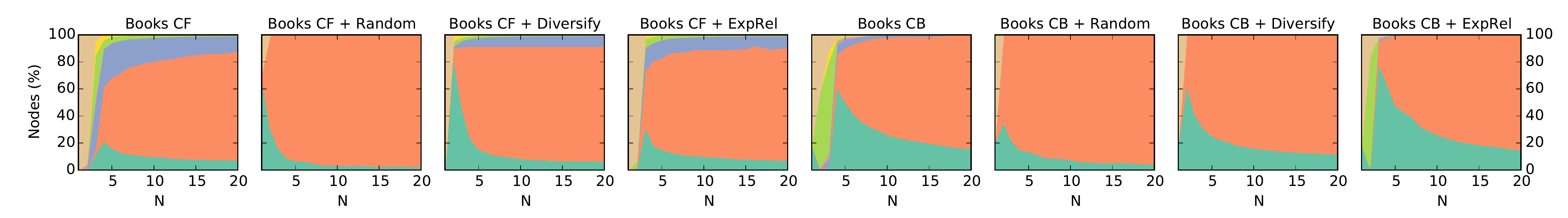}}
  \\[-2em] 
  \includegraphics[width=0.65\textwidth]{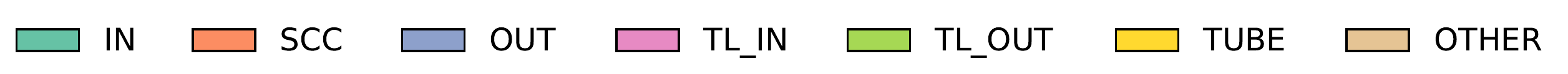}  
    \caption{\textbf{Bow tie analysis of the networks.} This figure plots the distributions of the component memberships according to the bow tie Model in the recommendations networks over $N$ (the number of recommendations) for MovieLens (a) and BookCrossing (b). For small $N$s, node memberships vary. With increasing $N$, most of the nodes are in the largest strongly connected component (SCC).}
  \label{fig:topology_bowtie_stacked}
  \savespace
\end{figure*}

The application of the three diversification algorithms to improve reachability overall had a positive effect. Replacing one recommendation with a random one led to a large increase in the size of the largest strongly connected component and to a largest component of $80$\% for 3 recommendations. As random connections are known to strongly increase reachability, we took this approach as our baseline. Both the \emph{Diversify} and the \emph{ExpRel} approaches led to an increase in the size of the largest component, with \emph{Diversify} performing better. For top-5 recommenders, introducting one diverse (but still relevant) one leads to a largest component comprising 60-80\% of all nodes, thus strongly improving reachability.

The clustering coefficient correlated negatively with the size of the largest component: the smaller the largest component, the higher the clustering coefficient--indicating the existence of isolated but highly clustered ``caves'' of items. When all links of a set of nodes are confined within a smaller component, that component is necessarily clustered more strongly. This might make navigation within the smaller component easier at the cost of not being able to reach larger parts of the network outside the component. The application of the three diversification algorithms led to lower clustering in the network. This indicates that while these algorithms connect more nodes to the core of the network, they might render navigability more difficult by removing some of the more obvious connections.

\paragraphcustom{Findings} We combine a set of global (component sizes) and local (clustering) measures to assess reachability of recommendation networks. For our datasets, recommendation networks are not well-connected, with between 20-40\% of nodes residing outside of the giant component in hundreds of small disconnected components. However, reachability can be improved by replacing the least relevant recommendation for each node with a diversified one. Clustering measures indicate that there exists a trade-off between stronger clustering (potentially facilitating navigation) and reachable parts of the network (where disconnectedness thwarts navigation).

\subsection{Efficient Reachability}

\paragraphcustom{Description} As the second step in assessing reachability, we investigated how \emph{efficiently} recommendation networks are reachable.
In order to obtain insight into the lengths of the shortest paths, we examined the eccentricity distributions of the largest components. The eccentricity of a node $i$ is the longest shortest path from $i$ to any other node in the component. This provided us with a means of gauging navigability from the path lengths.

\paragraphcustom{Results and Interpretation} Figure \ref{fig:topology} plots the distribution of the eccentricity values of all nodes in the largest components for $N=5$ recommendations (results for larger values of $N$ were qualitatively similar). Overall, we find that collaborative filtering led to shorter paths than content-based recommendations. This was in spite of collaborative filtering having larger strongly connected component--a phenomenon that has been observed in many types of graphs \cite{leskovec2005graphs}.

The diameters of the networks (i.e., the maximum eccentricity) for $N=5$ ranged from $14$ to $38$. Large diameters raise the question of whether users would actually undergo click sequences of this length to navigate the items. PageRank calculations \cite{brin} assume a teleporation factor of 15\%, meaning that in 15\% of clicks, a user does not follow a link on a web page but \emph{teleports} by manually typing in a new address or using a search engine. If we follow this model, then after five clicks the number of users following links within a recommender system has reduced to $45$\% and further decreases to $20$\% after $10$ clicks and $4$\% after $20$ clicks. While these are average estimates for the general Web, they clearly point out that standard recommender systems are not sufficiently navigable for $N=5$ recommendations. Furthermore these eccentricity values represent only lower bounds, as they would require users to always proceed on the shortest possible paths. Analysis of Wiki game data, where players actively try to find shortest paths, has shown that humans need at least two clicks more on average \cite{west2012human}.

A simple method to shrink the diameters is to increase the number of recommendations shown. And indeed we find that for increasing values of $N$, the eccentricity distributions shift towards smaller diameters. For $N=20$, most nodes in the collaborative filtering networks are reachable within $10$ steps. However, $20$ recommendations are considerably more than the five to twelve recommendations that most real-world systems use and potentially clutter user interfaces and make finding the right recommendation more difficult.

We found that diversification approaches were a more suitable means for reducing path lengths. Substituting a single recommendation shifted the eccentricity distributions closer towards shorter paths, while keeping recommendations relevant and the number of recommendations constant. We found that the \emph{Diversify} approach led to the largest improvement in terms of efficient reachability.

\paragraphcustom{Findings} We examine the distributions of the longest shortest path between nodes in the largest components. For our datasets, we find that distances between nodes (up to $39$ hops) are too long for reasonably efficient navigation. Collaborative filtering led to shorter paths than content-based recommendations. Diversification approaches prove to be a promising means to reducing path lengths.

\subsection{Partition Reachability}
\paragraphcustom{Description} After the analysis of effective and efficient reachability, we can conclude that recommendation networks lack reachability in many cases. As the next step, we now investigate the reachability of different sections in these networks. This analysis can give us more clarity as to what areas of the network are connected, and reveal one-way connections between parts of a recommender system.

\begin{figure}[t!]
    \centering
    \includegraphics[width=0.8\linewidth]{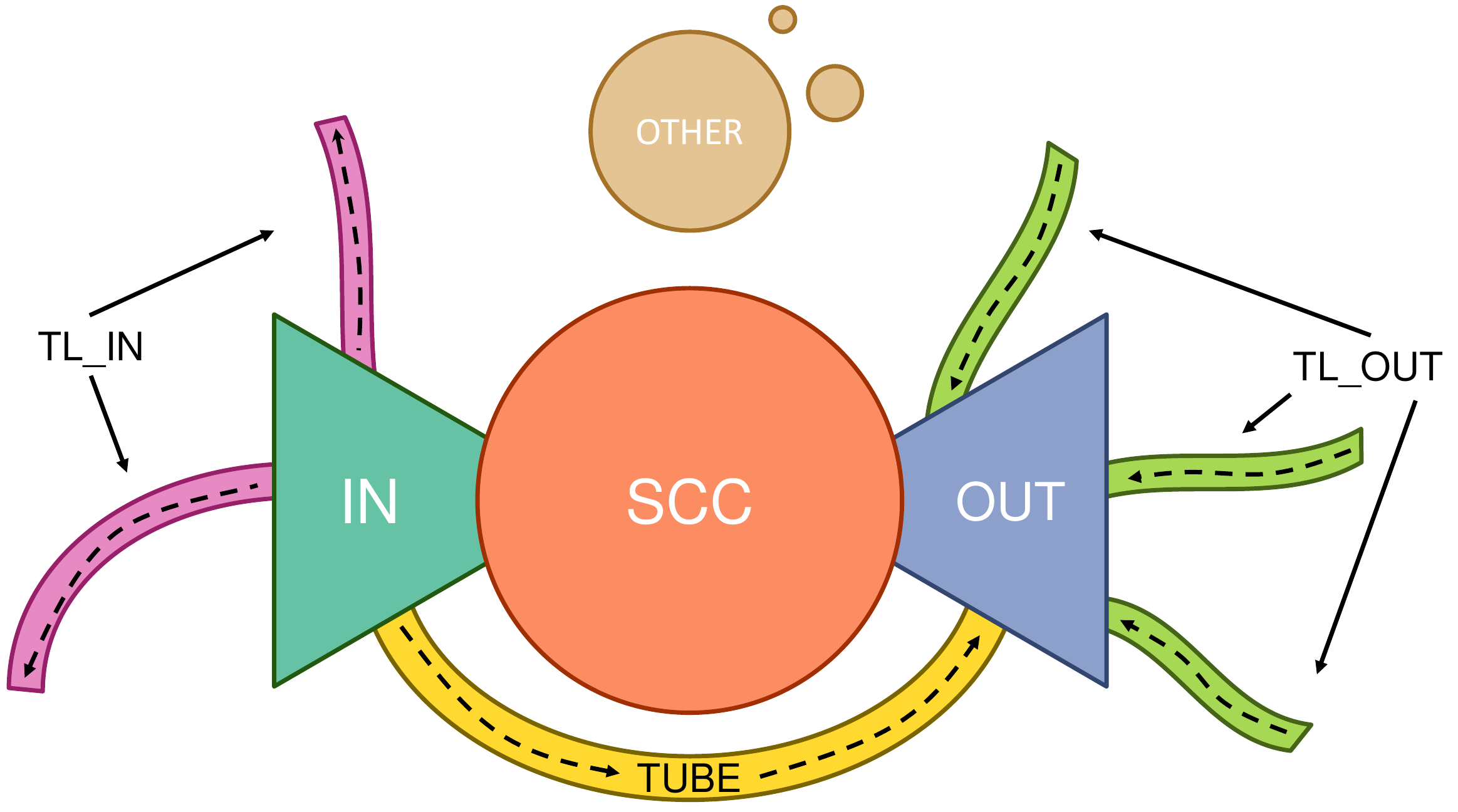}
    \caption{\textbf{Bow Tie Model.} The bow tie model \cite{bow_tie} partitions a network into a strongly connected component or core (\emph{SCC}), flanked by \emph{IN}, where nodes can reach the core but are not reachable from it and \emph{OUT}, where nodes are reachable from the core but not vice versa. Further components are \emph{TUBE}, providing an alternative route from \emph{IN} to \emph{OUT} and the \emph{TENDRILS (TL\_IN, TL\_OUT)} which contain nodes connected to \emph{IN} and \emph{OUT} which cannot reach the core. Remaining nodes are collected in \emph{OTHER}.
}
\label{fig:bowtie}
\savespace
\end{figure}

A prominent model for graph partitioning is the bow tie model \cite{bow_tie}, developed for the analysis of the Web. This model partitions a directed network into three major components: the largest strongly connected component (\emph{SCC}), wherein all nodes are mutually reachable, a component of all nodes from which \emph{SCC} can be reached (\emph{IN}) and a component of all nodes  reachable from \emph{SCC} (\emph{OUT}). Figure \ref{fig:bowtie} shows the model in more details and explains the components.

\paragraphcustom{Results and Interpretation} Figure \ref{fig:topology_bowtie_stacked} shows the partitioning of the recommendation networks by the bow-tie model. While for a small number of recommendations (i.e., low $N$) component division is relatively diverse, after about five recommendations all networks are partitioned mainly into \emph{IN} and \emph{SCC}. This implies that the network mainly consists of a strongly connected core and a partition leading to it.

A detailed analysis of where links from \emph{IN} component pointed to underlined this intuition: In all networks, more than two thirds of all links from items in \emph{IN} pointed to the \emph{SCC}. From a navigational perspective, this means that items in the \emph{SCC} can be reached from any item, but items in \emph{IN} are in most cases only reachable by direct selection, e.g., via search results. Note that with increasing $N$, nodes are bound to end up in the \emph{SCC} as density increases.

For collaborative filtering networks, the \emph{SCC} components were larger than for the content-based networks. This is likely due to the different features used in the recommendation generation. Collaborative filtering relied on ratings and connected items that were rated similarly, whereas content-based recommendations connected items based on their textual similarity. As such, content-based similarity tended to connect items that are described with the same key words. For instance, \emph{James Bond} movies mostly linked to movies of the same franchise for content-based recommendations. By contrast, for collaborative filtering, recommendations from \emph{James Bond} movies were more diverse and featured some recommendations to other action movies. Thus, the content-based approach was more strongly clustered and led to a smaller core that was harder to reach--a fact also visible from the higher clustering coefficients for content-based networks (cf. Figure \ref{fig:topology}).

The diversification approaches showed two effects:\begin{inparaenum}[(i)]
  \item they increased the size of \emph{SCC} and decreased the size of \emph{IN} and
  \item they strongly reduced the sizes of the rest of the components
\end{inparaenum}.
In terms of reachability, this is a desirable effect, as it makes the core (\emph{SCC}) reachable from a larger fraction of the items in the recommendation network.

The collaborative filtering networks for BookCrossing included a relevant number of nodes in the out-component (\emph{OUT}) and the out-tendril (\emph{TL\_OUT}) of the network for $N \geq 5$. Figure \ref{fig:topology_bowtie_alluvial} shows a membership change analysis for two selected networks. While initially (for $N=1$), a significant portion of the network is present in the form of the out-tendril and the out-component, nodes from these components then pass into the core (\emph{SCC}) with increasing $N$. For a smaller number of recommendations $N$ two separate strongly connected components with different sizes exist: \emph{SCC} and \emph{OUT}. With an increasing number of recommendations, \emph{SCC} attracted more items than \emph{OUT}. In addition, recommendations from some of the items in the \emph{SCC} pointed to \emph{OUT} items (but not vice versa), which connected the two components. An explanation for this situation could be the average number of ratings for items, which was significantly higher for \emph{SCC} items and lower for items in the \emph{OUT} component. As recommendations were calculated based on cosine similarity, items with few co-ratings were more likely to reciprocate their recommendations for other items with only few ratings, and popular items with many co-ratings were more likely to recommend other popular items. This made items in \emph{OUT} more likely to remain in the component in case of collaborative filtering (the problem with co-ratings was not present for content-based recommendations).

\paragraphcustom{Findings} We analyze recommendation networks based on the bow tie model, which partitions networks into components based on reachability. We find that the networks consist mainly of a strongly connected core of popular items together with an \emph{IN} component leading to the core. This implies that the core is reachable from most items. With diversified recommendations, networks have more components in the core and the \emph{IN} components, thus making the network better connected. Constructing navigable recommender systems could potentially be facilitated with the help of a modified similarity measure which is less harsh on the number of total ratings per item. The bow-tie model could then be used to evaluate and select appropriate similarity measures.

\newlength{\w}
\setlength{\w}{3.27em}
\begin{figure}[t!]
  \centering
  \raisebox{-0.7ex}{\scriptsize$N = 1$}\hspace{\w}\raisebox{-0.7ex}{\scriptsize$N=5$}\hspace{\w}\raisebox{-0.7ex}{\scriptsize$N = 10$}\hspace{\w}\raisebox{-0.7ex}{\scriptsize$N=15$}\hspace{\w}\raisebox{-0.7ex}{\scriptsize$N=20$}%
  
  \includegraphics[width=\linewidth]{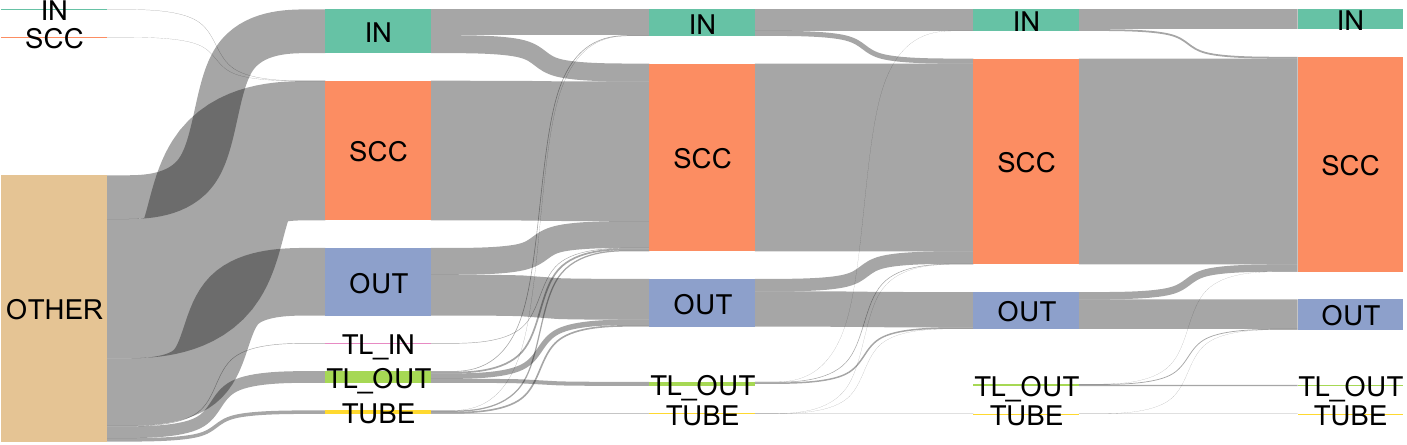}
    \caption{\textbf{Membership change analysis of the bow tie structure of the collaborative filtering BookCrossing network.} The figure plots the node membership changes between selected values of $N$ (the number of recommendations). While initially most nodes belong to the \emph{OTHER} component, nodes move to \emph{IN} and \emph{SCC} with increasing $N$. As a particularity for this algorithm and dataset, the \emph{OUT} component persists with increasing $N$. This is due to node intake from  \emph{TL\_OUT}.}
  \label{fig:topology_bowtie_alluvial}
  \savespace
\end{figure}

\section{Navigability}
\label{sec:navigability}
\begin{figure*}[t!]
    \centering
    \subfloat[Point-To-Point Search]{\includegraphics[page=1,width=0.225\linewidth]{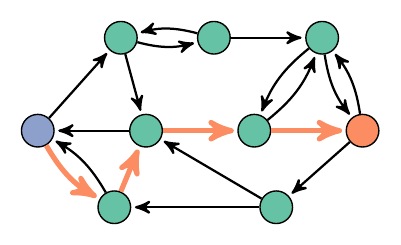}}
    \subfloat[Berrypicking]{\includegraphics[page=2,width=0.225\linewidth]{scenarios.pdf}}
    \subfloat[Information Foraging]{\includegraphics[page=3,width=0.225\linewidth]{scenarios.pdf}}    
    \subfloat[]{\includegraphics[page=4,width=0.225\linewidth]{scenarios.pdf}}    
  \vspace{-0.5em}
  \caption{\textbf{Information Seeking Scenarios.} This figures shows the three information seeking scenarios used in our analysis. The goal in Point-To-Point Search was to find a single target node. For Berrypicking, we clustered the networks and set the goal of finding any one node in four predetermined clusters (shown in gray). For Information Foraging, the goal was to find all nodes in one predetermined cluster.}
\savespace
\label{fig:scenarios}
\end{figure*}

In the first part of this article, we investigated the reachability of recommender systems. As the second step of our analysis, we now focus our attention on the dynamics of searching and navigating recommendation networks. In a typical information seeking model, users move from one item to another by traversing recommendations. This activity can be intertwined with using the search function--e.g., exploring the results, backtracking and taking another path through the recommendations, or simply entering a refined search query in the search field \cite{white}. Browsing is an important use case of a recommender system, as many users find browsing pleasant \cite{herlocker2004evaluating} or use it to discover new content \cite{lerman2007social}. A defining property of this process is that the knowledge users have about the network is generally local: users only know about the links emanating from the current node and have intuitions about where those links lead.

Several information seeking models have been established in the literature to model navigation and exploration in information networks. In this paper we concentrate on three of these models: 
\begin{itemize}
\itemsep0em 
\item Point-to-point Search \cite{kleinberg2000},
\item Berrypicking \cite{bates1989design}
\item Information Foraging \cite{pirolli2007}
\end{itemize}
We examine the navigability of recommendation networks by simulating scenarios based on these three information seeking models and by measuring the success of the simulations in achieving a given navigation goal. This analysis shows us to what extent recommendation networks are suitable for dynamic processes such as navigation and exploration. This evaluation goes beyond a standard one-click evaluation scenario in recommender systems--it is in particular an inspection of the suitability of these networks to accommodate users in following several sequential recommendations, one after the other.

For all simulations, we applied a \emph{greedy search} mechanism. We assumed that a function existed for each node which we could evaluate for each outgoing link in reference to the current navigation goal. The simulation then always acted greedily and selected the link maximizing this function (or backtracked in case of a dead end). We ran simulations for a total of $50$ steps per goal. In particular, the next node was selected greedily based on a background knowledge, which we represented as a matrix $S$, containing similarities between pairs of nodes (i.e., items). For a pair of nodes $(i, j)$, where $i$ is a candidate node and $j$ is the target node, $S_{ij}$ represented the function value for the candidate node $i$. We used the following types of background knowledge:
\begin{enumerate}[itemsep=2pt,parsep=0pt,leftmargin=*,label=(\roman*)]
    \item \emph{Title}: The similarity matrix $S$ contained the cosine similarity of the TF-IDF features for item titles. This represented the intuitions about navigation gained from looking at the titles of recommendation targets.
    \item \emph{Neighbors}: $S$ held the cosine similarity of the vector of neighbors of nodes in the recommendation network. This represented intuitions about areas of the network--e.g., a science fiction film likely leads to more films of the same genre.
    \item \emph{Wikipedia neighbors}: $S$ contained the cosine similarity of the vector of neighbors of nodes in the Wikipedia network (with items mapped to Wikipedia articles). This represented similar intuitions as those for the \emph{Neighbors} background knowledge, but with the information originating from an external body of knowledge.
    \item \emph{Shortest Paths (Optimal Solution)}: For the optimal solution, we used the matrix containing information on the shortest paths.
    \item \emph{Random}: This matrix consisted of all zeros (containing no background knowledge at all) which resulted in a random walk.    
\end{enumerate}

We believe that these different types of knowledge represent a good inventory of possible intuitions for navigating recommendation networks. We found that the best-performing background knowledge leads to success ratios well within the baselines (the random walk and the shortest paths) in all cases. For the sake of brevity, we only report the best-performing approaches for each network in the following section.

\subsection{Point-To-Point Search}

\begin{figure}[t!]
\centering
\includegraphics[width=0.8\linewidth]{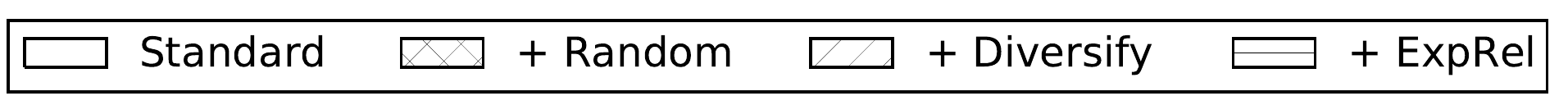}
\\[-1em]
\subfloat[MovieLens]{\includegraphics[width=0.4\linewidth]{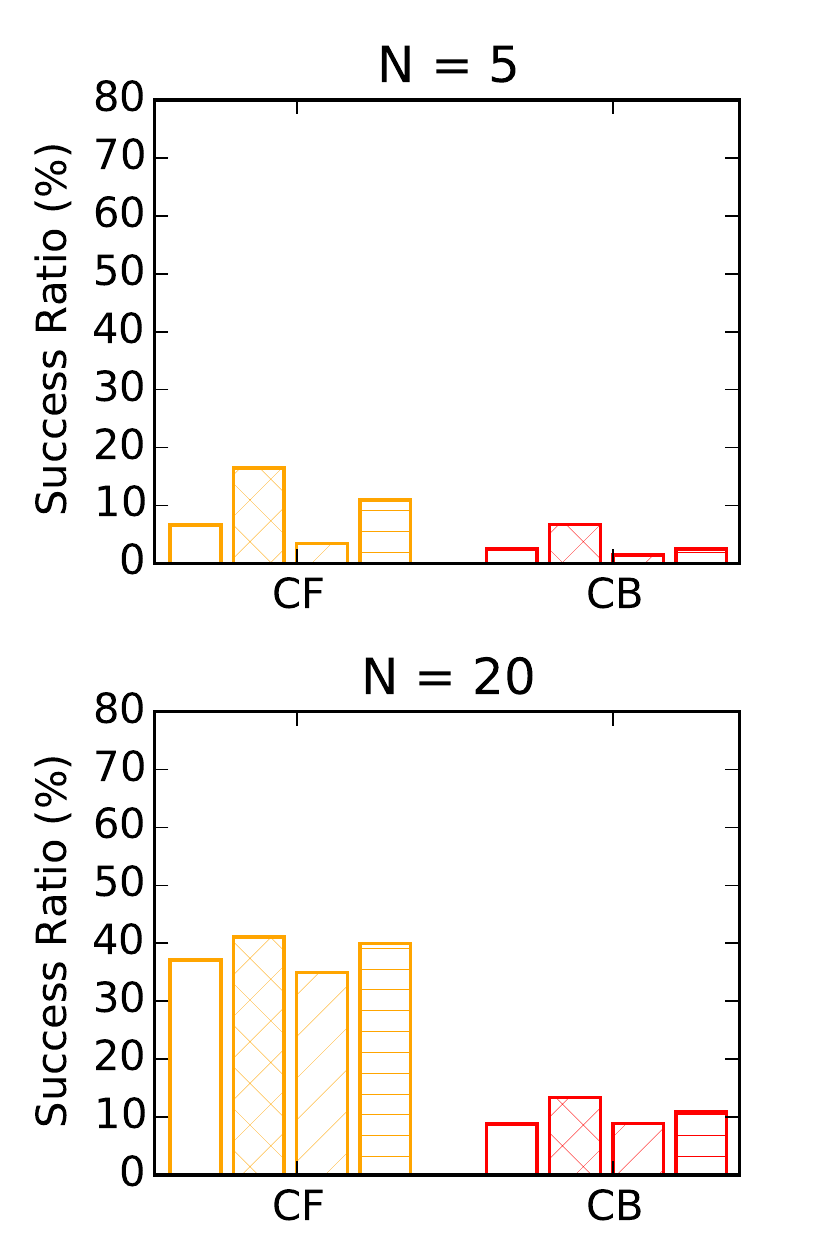}}
\subfloat[BookCrossing]{\includegraphics[width=0.4\linewidth]{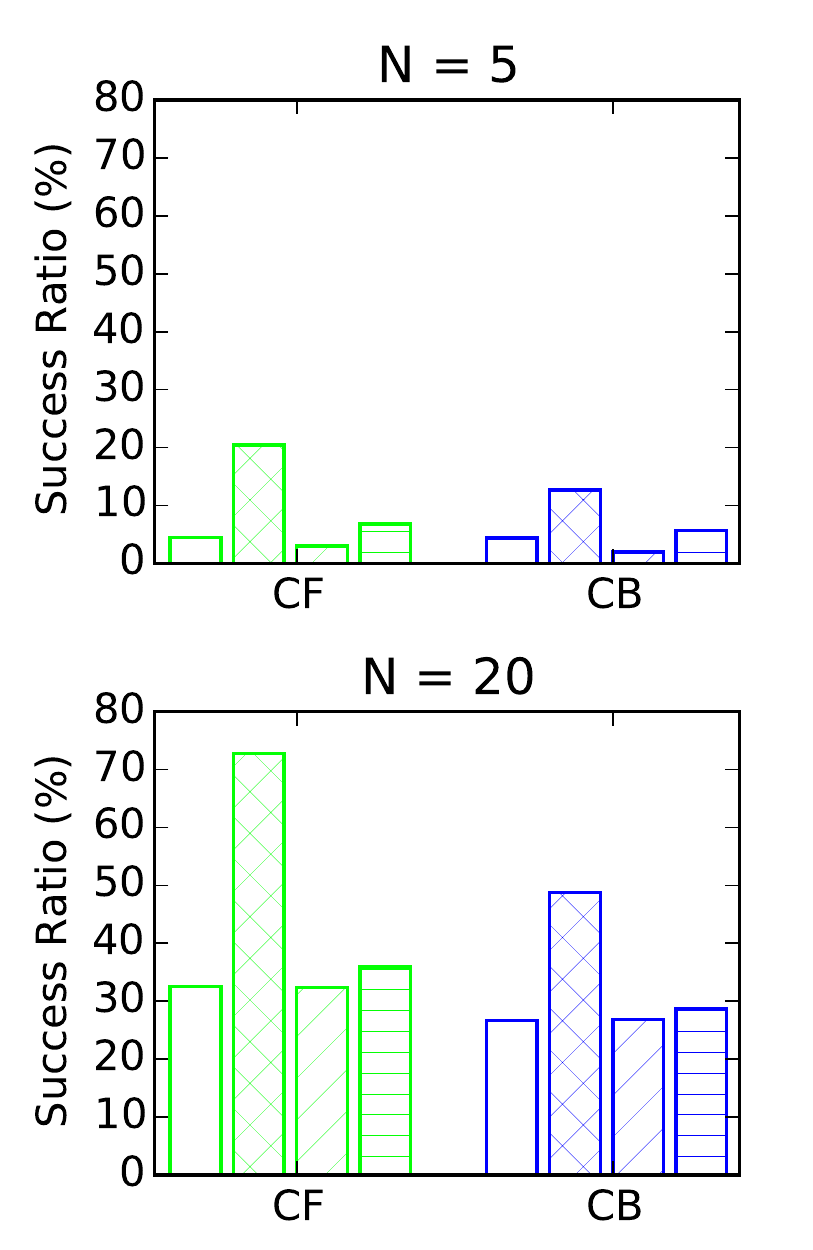}}
\vspace{-0.5em}
\caption{\textbf{Success ratios for the Point-To-Point search scenario for $N=5$ and $N=20$ recommendations.} Success ratios were better for a larger number of recommendations and were also improved by diversification. Collaborative Filtering (CF) led to better results than Content-Based methods (CB).}
\label{fig:navigation:point-to-point}
\savespace
\end{figure}

\paragraphcustom{Description} Point-To-Point search represents the task of finding a single target item in the recommendation network. We randomly sampled $1,200$ pairs of nodes from each  network, not taking reachability into account. We then ran navigation simulations for all of these pairs, starting at the start node of the pair and with the objective of reaching the target node. As an example, in simulations with \emph{Title} background knowledge, the next node to go to was always (greedily) chosen to be the neighboring node with the most similar title to the target. Note that the outcome was therefore affected by both the reachability and navigability of the network. Figure \ref{fig:scenarios} displays an example of a Point-To-Point scenario.

\paragraphcustom{Results and Interpretation} Figure \ref{fig:navigation:point-to-point} displays the success ratio (i.e., the fraction of successful simulations), showing only the result with the best-performing background knowledge. In the case of MovieLens, this was the \emph{Title} background knowledge and for BookCrossing the \emph{Neighbors} knowledge. This is likely an artifact of the higher clustering in the BookCrossing networks, suggesting that these networks were better suited to guiding search with intuitions about the common \emph{Neighbors} towards the target. In the case of MovieLens by contrast, the \emph{Title} similarities proved to be better in achieving this, indicating that film titles in our datasets were more indicative of the general area in the network than book titles. The same ranking of background knowledge was also the case for the Berrypicking and the Information Foraging scenarios.

Overall, performance with Point-To-Point search was not very satisfactory for most of the recommendation networks investigated. For $N=5$ recommendations, only 2-6\% of targets were found for standard recommendation algorithms, which was increased by up to 20\% with diversification approaches. Simulations on collaborative filtering networks generally outperformed those on the content-based networks. With increased number of recommendations we observed a significant performance gain. For $N=20$ recommendations,  8-40\% of targets were found for standard algorithms, and up to 70\% for diversified approaches. As in the analysis of reachability, we took the random diversification as an upper bound of the possible increase when substituting a single recommendation. In contrast to reachability, for navigability the \emph{Diversify} approach actually performed slightly worse than standard algorithms, while \emph{ExpRel} led to an increase. This suggests that a trade-off exists between improving reachability and navigability.

\paragraphcustom{Findings}
In evaluating the Point-To-Point Search scenario, we find that the networks are not well suited to this navigation model using standard algorithms. This situation can be somewhat improved by increasing the number of recommendations and by applying diversification algorithms.

\subsection{Berrypicking}

\paragraphcustom{Description} Berrypicking is an information seeking model proposed by Marcia J. Bates \cite{bates1989design}, which regards information seeking as a dynamic and evolving process. By contrast, Point-To-Point Search might be regarded as a static model of information seeking where the information need remains constant throughout a complete navigation session. In Berrypicking, the information need is evolving and can be satisfied by multiple pieces of information in a \emph{bit-at-a-time retrieval} \cite{bates1989design}--an analogy to picking berries on bushes, where berries are scattered and must be picked one by one.

Based on Berrypicking, we evaluated the following navigation scenario in our recommendation networks: We first created clusters of network nodes based on publication year and genres. We then chose all clusters with 3 to 30 nodes as input and randomly sampled a total of $1,200$ subsets containing four cluster each. With these as input, we randomly choose one node from the first cluster as the starting point in the network. The objective of the scenario was then to reach an arbitrary node from the second cluster, followed by an arbitrary node from the third and finally an arbitrary node from the forth cluster. In this way, the scenario modeled the evolving stages of Berrypicking with information needs changing after every discovered item.

\begin{figure}[t!]
    \centering
    \includegraphics[width=0.8\linewidth]{nav_legend.pdf}
    \\[-1em]
\subfloat[MovieLens]{\includegraphics[width=0.4\linewidth]{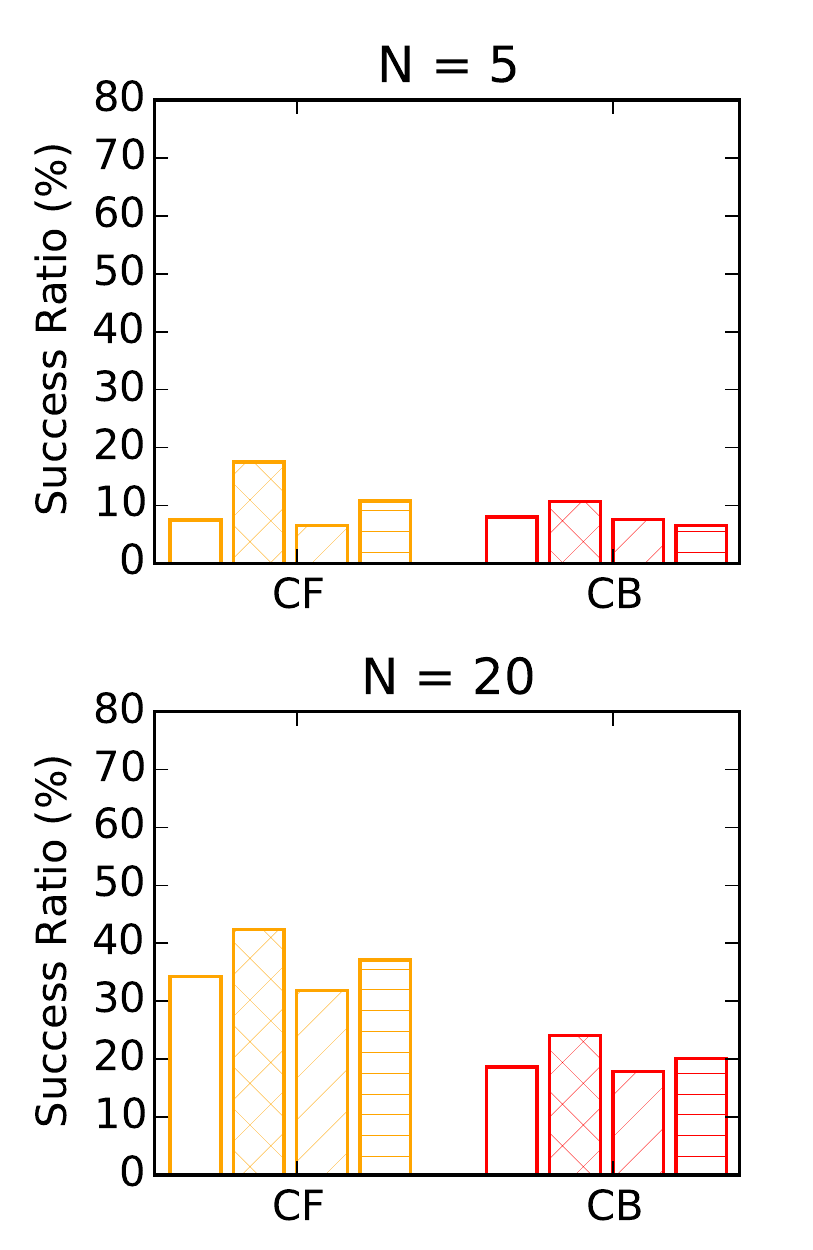}}
\subfloat[BookCrossing]{\includegraphics[width=0.4\linewidth]{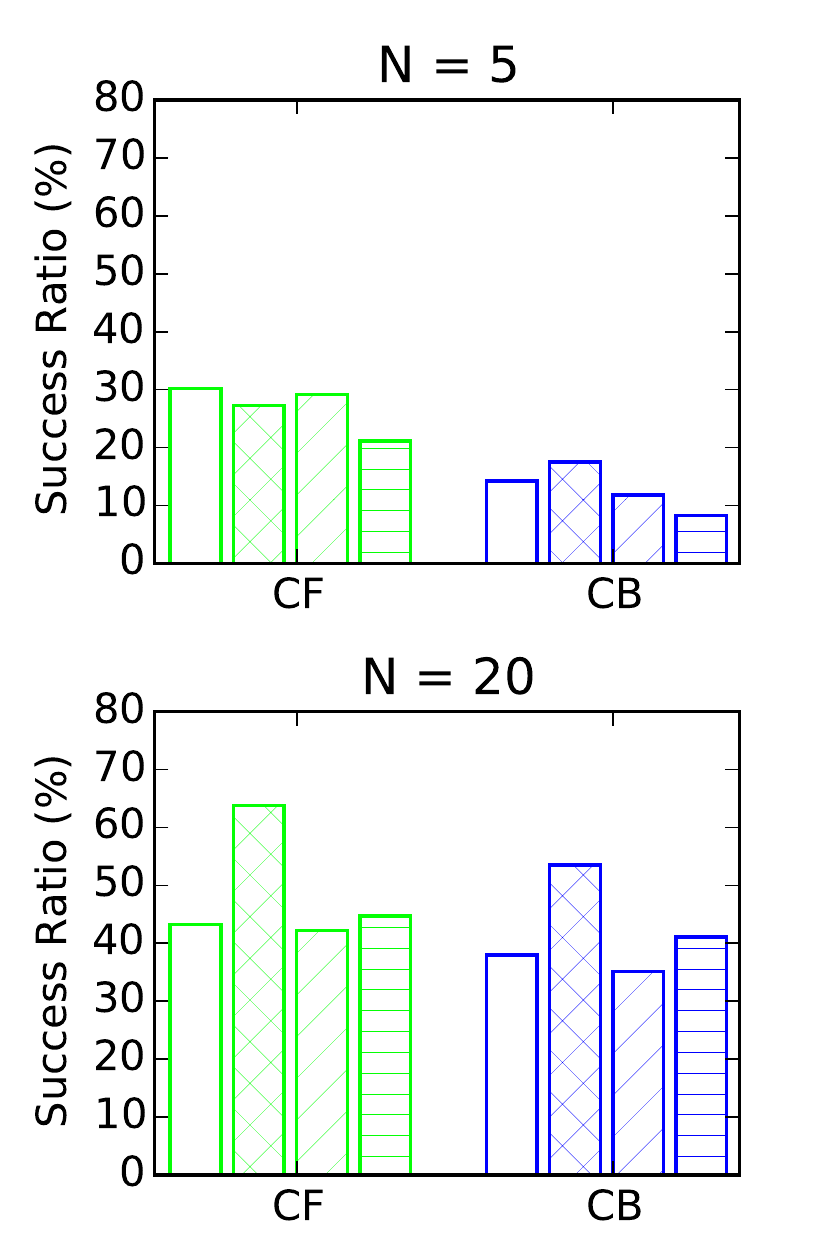}}    
  \vspace{-0.5em}
  \caption{\textbf{Success ratios for the Berrypicking scenario for $N=5$ and $N=20$ recommendations.} Success ratios were better for a larger number of recommendations and were also improved by diversification. Collaborative Filtering (CF) led to better results than Content-Based methods (CB).}
\label{fig:navigation_berrypicking}
\savespace
\end{figure}

The basic implementation of this scenario was the same as for Point-To-Point Search, using the same three types of background knowledge. A difference here,, however, was that the target of the navigation was now not a single node but the cluster centroid (using the average similarity of all nodes in the cluster). The next node to go to was then chosen as the node with the highest similarity to the centroid in every step.

\paragraphcustom{Results and Interpretation} Similar observations as those for the Point-To-Point Search can be made for the Berrypicking scenario: for a small number of recommendations, none of the networks performed well. In this case, the success ratio was the average percentage of targets found: 10\% success for Berrypicking means that, on average, 10\% of three targets (instead of one in the case of Point-To-Point Search) had been found. This implies that even though the success ratio was almost the same as for Point-To-Point Search more targets were found overall. This suggests that the recommendation networks we studied were better suited to supporting Berrypicking than Point-to-Point Search.

Even with diversification approaches applied and $N=20$ recommendations, however, only 20-40\% of scenarios were successfully completed. This indicates that while one or two clusters were found in the simulations, finding all clusters proved to be too difficult. For recommender systems, the combination of recommendations with an efficient search function is therefore vital to support information seeking and browsing.

\paragraphcustom{Findings} We find that for Berrypicking, a scenario representing dynamic information search, was somewhat better supported than Point-to-Point Search. A high number of recommendations and diversification led to success ratios around 40\%.

\subsection{Information Foraging}

\paragraphcustom{Description} Information Foraging \cite{pirolli2007} is an information seeking theory inspired by Optimal Foraging Theory in nature, where organisms have adopted strategies maximizing energy intake per time unit. For instance, when foraging on a patch of food (e.g., apples on a tree), animals must decide when to move on to the next patch (e.g., if finding new apples on the current tree has become too strenuous or all apples have been consumed). In the 1990s, Peter Pirolli and others found that some of the same mechanisms hold good for human information seeking behavior, and that humans try to maximize the information gain per time unit. Information is perceived as occurring in \emph{patches}, indicated by \emph{information scent} \cite{chi2001using}.

In a scenario based on Information Foraging, we model the scenario of depleting a patch of information in a recommendation network. We assume that a after using the search function and arriving at one of the nodes in an information patch (i.e., a cluster as defined in Berrypicking), the objective is now to find all the other nodes in the patch--guided by information scent in terms of the background knowledge, which represents intuitions about items.

The implementation of this scenario was very similar to the one used for Berrypicking: Navigation was directed not at single nodes but towards a cluster centroid.

\paragraphcustom{Results and Interpretation} The simulations for the Information Foraging scenario performed best in the MovieLens collaborative filtering network. This suggests that the clustering (which was performed based on year of publication and genres) was best represented in this network. Furthermore, titles were generally more indicative of targets in this network, with \emph{Title} being the best-performing background knowledge. While one could expect that this method of clustering would favor the content-based networks (as title and categories are part of the textual content), this was evidently not the case. Another possible reason for this behavior could be the number of components and the size of the largest connected component: the MovieLens collaborative-filtering network had the largest giant component and the smallest number of components. Evidently, if the task was to visit as many nodes as possible from a specific part of the network, the reachbility of that part of the network plays the
most important role. In general, recommendation networks seem to suffer from the existence of large number of (almost) isolated ``caves'' that are only loosely connected to each other. In the MovieLens collaborative filtering network, this problem was to some extent solved by a larger number of recommendation links that better connect the ``caves'' and formed a larger connected component. This analysis shows that, apart from shortening the shortest paths in the recommendation networks, injecting more links may improve navigability by ensuring a more robust connectedness of the item ``caves'' and the networks in general.

The success ratio for this scenario was again based on the total number of target nodes. With the best approaches finding 30-40\% of nodes in 10 steps, this might already be to some extent satisfying for recommendation networks in exploratory scenarios. Diversification approaches showed only a small improvement in terms of success ratio. One possible explanation could be that the clustering coefficients, which were negatively affected by diversification, had a substantial influence on the results. A diversification algorithm specifically connecting caves or clusters in the network could improve the results here, but is beyond the scope of this analysis.

\begin{figure}[t!]
    \centering
    \includegraphics[width=0.8\linewidth]{nav_legend.pdf}
    \\[-1em]
\subfloat[MovieLens]{\includegraphics[width=0.4\linewidth]{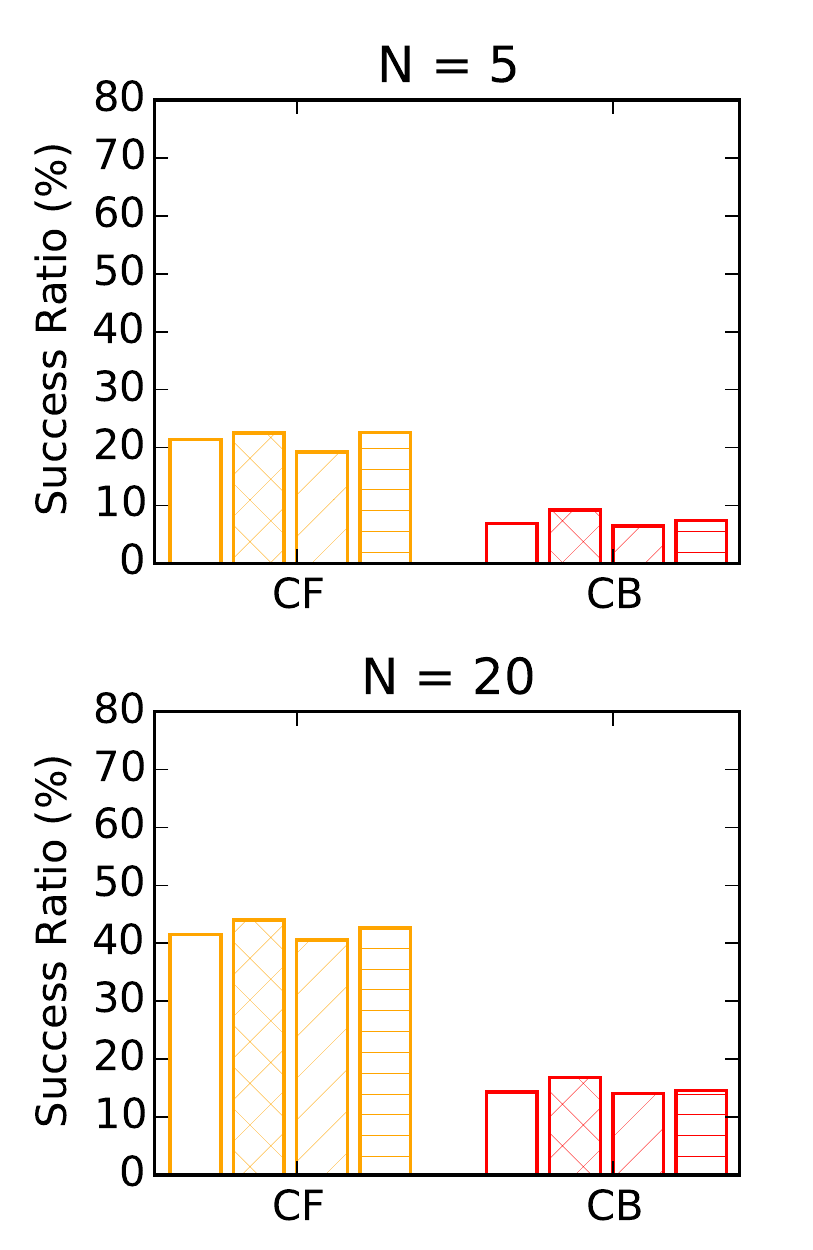}}
\subfloat[BookCrossing]{\includegraphics[width=0.4\linewidth]{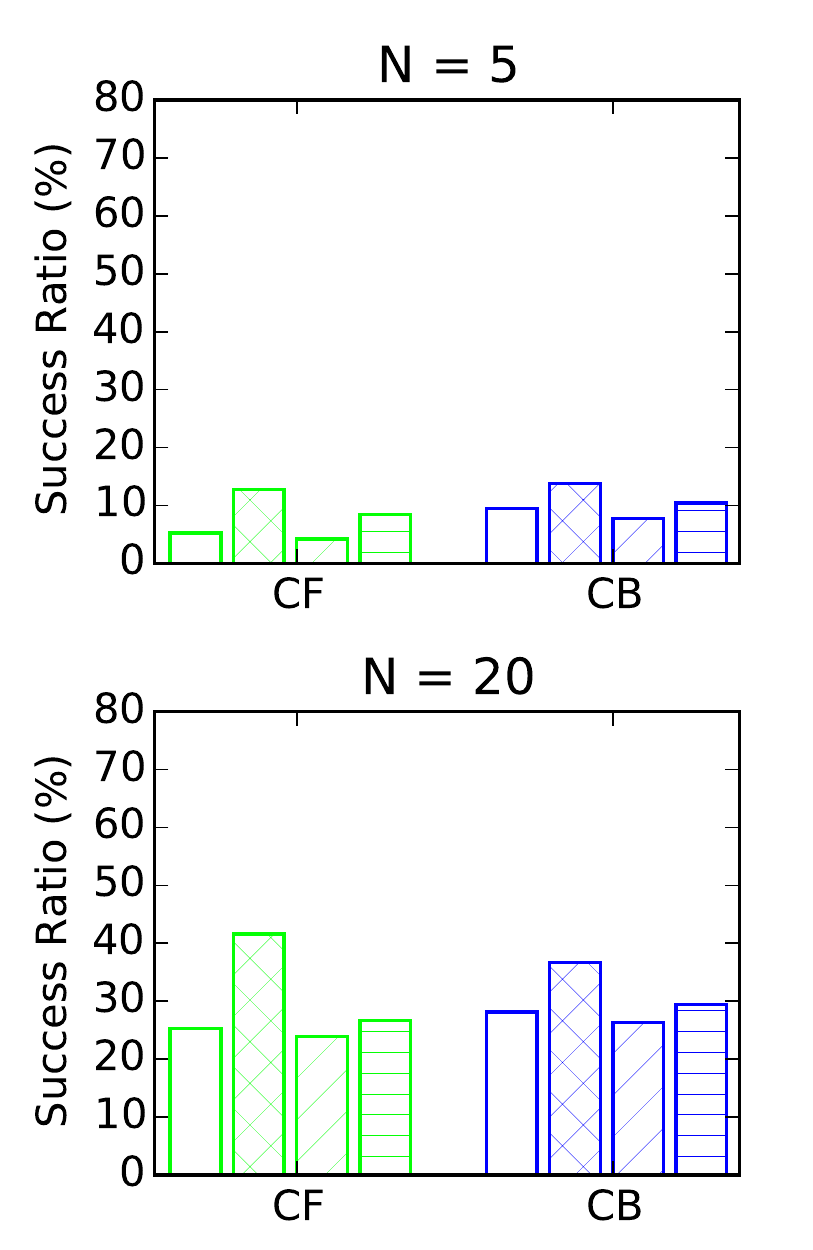}}     
  \vspace{-0.5em}
  \caption{\textbf{Success ratios for the Information Foraging Search scenario for $N=5$ and $N=20$ recommendations.} Success ratios were better for a larger number of recommendations and were also improved by diversification. Collaborative Filtering (CF) led to better results than Content-Based methods (CB).}
\label{fig:navigation_information_foraging}
\savespace
\end{figure}

\paragraphcustom{Findings} We find that for an explorative scenario based on Information Foraging, similar outcomes as for the other scenarios hold true: it was not supported very well in our datasets. This analysis reveals another important structural deficiency of recommender networks: poor connectivity between item ``caves''. In order to improve navigability of recommendation networks this deficiency could be overcome by specifically connecting components.
\section{Discussion}
\label{sec:discussion}
In this work, we evaluated recommendation networks in the context of reachability and navigation dynamics. We presented a general approach and applied it to two networks. We found that recommendation networks created with standard recommendation algorithms did not fare well in terms of reachability nor navigability, and showed that this could be improved with diversification approaches.

The navigation models applied in this approach are largely well-established in the research community and cover a wide range of typical user interaction scenarios with information systems in general, and recommender systems in particular. Greedy search, the basis for our navigation scenarios based on these models, has been used in previous work to analyze navigation dynamics in networks \cite{helic2013models, lamprecht2015using}. The navigation models were deliberately kept simple, as the focus of our work was not on the information seeking models and their validity but on the \textit{properties of recommendation networks}. Our evaluation approach does not depend on a particular model, which can be adapted or exchanged in future work. Possible enhancements include teleportation \cite{brin} to model the interplay with search function, stochastic instead of deterministic action selection, or a learning component, e.g., for memorizing preferred paths.

The collaborative filtering approach used in the generation of the networks is very basic. Expanding this work to personalized recommendations would represent a logical expansion to our work. It must be noted, however, that this would also lead to distinct networks for each and every user and would require larger-scale analysis. In using non-personalized collaborative filtering, we effectively inspect recommendation networks for users who are new to the system or simply browsing the page without being registered--having assigned no ratings, the system can only suggest the globally most similar items.

Additional future work could include content-based recommendations based on more elaborate features. As of now, collaborative filtering features appear to lead to a better comprehension of items, as suggested by the topology and navigation results. In future work, other, more sophisticated text feature algorithms such as LSA or LDA could be used to possibly improve on this. However, our evaluation approach is general enough to accommodate all these future extensions.

We investigated three distinct diversification algorithms as possible improvements to reachability and navigability. In the analysis of reachability, we found that the \emph{Diversify} method performed best and was close to the addition of a random recommendation, while at the same time ensuring the relevance of the included recommendation. By contrast, for navigability \emph{Diversify} mostly led to a slight decrease in success, whereas the \emph{ExpRel} approach was able to improve results. This results point to differences between reachability and navigability of recommender systems. A more detailed investigation will be necessary in the future to combine the best elements of both approaches and develop a diversification measure supporting both reachability and navigability.

Another approach to improving navigability would be to increase the number of recommendations shown. However, this risks cluttering the user interface with too many recommendations and is therefore mostly avoided in real-world recommender systems. An alternative could be to keep the number of recommendations constant but to add more diversification and evaluate the fraction of diversity and navigability that users are willing to trade in for accuracy.

\section{Conclusions}
\label{sec:conclusions}

We have presented a general approach to evaluating the reachability and navigability of arbitrary recommendation networks. Our approach is based on an evaluation conducted on two levels: First we evaluate the topology of recommendation networks by looking at components, eccentricity and bow-tie structure. Second, we evaluate the dynamics of recommendation networks by simulating three different navigation models, namely Point-To-Point Search, Information Foraging and Berrypicking. We applied this approach to two datasets and found that reachability and navigability was not well-supported for standard recommendation algorithms. None of the recommendation networks was navigable in any of the scenarios, if plausible and practical constraints are applied. We investigated possible improvements for this and found that the results could be improved with simple diversification approaches.

We find that in our datasets, collaborative-filtering performed better than a content-based approach, suggesting that exploiting the collective knowledge present in ratings leads to more easily navigable recommender systems. While the results of our experiments are limited to the datasets under investigation, our approach to evaluating the navigability of recommendation networks is general. It can be applied to arbitrary recommendation networks, thereby acting as a novel tool of measurement for an increasingly important dimension of recommendation systems. We hope that our work stimulates more research on evaluating and ultimately improving the navigability of recommendation systems and corresponding algorithms.

\section{Acknowledgments}
This research was supported by a grant from the Austrian Science Fund (FWF) [P24866].
\vspace{-0.5em}
\balance
\bibliographystyle{abbrv}
\bibliography{bibliography} 

\begin{thebibliography}{10}

\bibitem{bates1989design}
M.~J. Bates.
\newblock The design of browsing and berrypicking techniques for the online
  search interface.
\newblock {\em Online Information Review}, 13(5):407--424, 1989.

\bibitem{boim2011diversification}
R.~Boim, T.~Milo, and S.~Novgorodov.
\newblock Diversification and refinement in collaborative filtering
  recommender.
\newblock In {\em CIKM'11}, pages 739--744, 2011.

\bibitem{brin}
S.~Brin and L.~Page.
\newblock The anatomy of a large-scale hypertextual web search engine.
\newblock In {\em {WWW '98}}. Elsevier Science Publishers B. V., 1998.

\bibitem{bow_tie}
A.~Broder, R.~Kumar, F.~Maghoul, P.~Raghavan, S.~Rajagopalan, R.~Stata,
  A.~Tomkins, and J.~Wiener.
\newblock Graph structure in the web.
\newblock {\em Computer networks}, 33(1):309–320, 2000.

\bibitem{cano_topology_2005}
P.~Cano, O.~Celma, M.~Koppenberger, and J.~M. Buldú.
\newblock Topology of music recommendation networks.
\newblock {\em Chaos: An Interdisciplinary Journal of Nonlinear Science},
  16(1), 2006.

\bibitem{celma_new_2008}
{\`O}.~Celma and P.~Herrera.
\newblock A new approach to evaluating novel recommendations.
\newblock In {\em {RecSys'08}}, 2008.

\bibitem{chi2001using}
E.~H. Chi, P.~Pirolli, K.~Chen, and J.~Pitkow.
\newblock Using information scent to model user information needs and actions
  and the web.
\newblock In {\em CHI'01}, pages 490--497, 2001.

\bibitem{davidson2010youtube}
J.~Davidson, B.~Liebald, J.~Liu, P.~Nandy, T.~Van~Vleet, U.~Gargi, S.~Gupta,
  Y.~He, M.~Lambert, B.~Livingston, et~al.
\newblock {The YouTube Video Recommendation System}.
\newblock In {\em RecSys'10}, pages 293--296. ACM, 2010.

\bibitem{helic2013models}
D.~Helic, M.~Strohmaier, M.~Granitzer, and R.~Scherer.
\newblock Models of human navigation in information networks based on
  decentralized search.
\newblock In {\em {HT'13}}, 2013.

\bibitem{herlocker2004evaluating}
J.~L. Herlocker, J.~A. Konstan, L.~G. Terveen, and J.~T. Riedl.
\newblock Evaluating collaborative filtering recommender systems.
\newblock {\em ACM Transactions on Information Systems (TOIS)}, 22(1):5--53,
  2004.

\bibitem{kleinberg2000a}
J.~Kleinberg.
\newblock The small-world phenomenon: an algorithm perspective.
\newblock In {\em {STOC '00}}, 2000.

\bibitem{kleinberg2001}
J.~Kleinberg.
\newblock Small-world phenomena and the dynamics of information.
\newblock In {\em {NIPS 14}}, Cambridge, MA, USA, 2001. MIT Press.

\bibitem{kleinberg2000}
J.~M. Kleinberg.
\newblock Navigation in a small world.
\newblock {\em Nature}, 406(6798):845, August 2000.

\bibitem{kuccuktuncc2013diversified}
O.~K{\"u}{\c{c}}{\"u}ktun{\c{c}}, E.~Saule, K.~Kaya, and {\"U}.~V.
  {\c{C}}ataly{\"u}rek.
\newblock Diversified recommendation on graphs: pitfalls, measures, and
  algorithms.
\newblock In {\em WWW'13}, pages 715--726, 2013.

\bibitem{lamprecht2015using}
D.~Lamprecht, M.~Strohmaier, D.~Helic, C.~Nyulas, T.~Tudorache, N.~F. Noy, and
  M.~A. Musen.
\newblock Using ontologies to model human navigation behavior in information
  networks: A study based on wikipedia.
\newblock {\em Semantic Web}, 2015.

\bibitem{lerman2007social}
K.~Lerman and L.~Jones.
\newblock {Social Browsing on Flickr}.
\newblock In {\em ICWSM'07}, 2007.

\bibitem{leskovec2005graphs}
J.~Leskovec, J.~Kleinberg, and C.~Faloutsos.
\newblock Graphs over time: densification laws, shrinking diameters and
  possible explanations.
\newblock In {\em KDD'05}, pages 177--187, 2005.

\bibitem{marchionini2006exploratory}
G.~Marchionini.
\newblock Exploratory search: from finding to understanding.
\newblock {\em Communications of the ACM}, 49(4):41--46, 2006.

\bibitem{milgram67}
S.~Milgram.
\newblock The small world problem.
\newblock {\em Psychology Today}, 1:60--67, 1967.

\bibitem{mirza2003studying}
B.~J. Mirza, B.~J. Keller, and N.~Ramakrishnan.
\newblock Studying recommendation algorithms by graph analysis.
\newblock {\em Journal of Intelligent Information Systems}, 20(2):131--160,
  2003.

\bibitem{nakatsuji2010classical}
M.~Nakatsuji, Y.~Fujiwara, A.~Tanaka, T.~Uchiyama, K.~Fujimura, and T.~Ishida.
\newblock Classical music for rock fans?: Novel recommendations for expanding
  user interests.
\newblock In {\em CIKM'10}, pages 949--958. ACM, 2010.

\bibitem{nguyen2014exploring}
T.~T. Nguyen, P.-M. Hui, F.~M. Harper, L.~Terveen, and J.~A. Konstan.
\newblock Exploring the filter bubble: the effect of using recommender systems
  on content diversity.
\newblock In {\em WWW'14}, pages 677--686, 2014.

\bibitem{pirolli2007}
P.~Pirolli.
\newblock {\em Information Foraging Theory: Adaptive Interaction with
  Information}.
\newblock Oxford Univ. Press, 2007.

\bibitem{resnick1997recommender}
P.~Resnick and H.~R. Varian.
\newblock Recommender systems.
\newblock {\em Communications of the ACM}, 40(3):56--58, 1997.

\bibitem{seyerlehner_limitations_2009}
K.~Seyerlehner, A.~Flexer, and G.~Widmer.
\newblock On the limitations of browsing {top-N} recommender systems.
\newblock In {\em {RecSys'09}}, 2009.

\bibitem{seyerlehner_browsing_2009}
K.~Seyerlehner, P.~Knees, D.~Schnitzer, and G.~Widmer.
\newblock Browsing music recommendation networks.
\newblock In {\em {ISMIR'09}}, 2009.

\bibitem{shani2011evaluating}
G.~Shani and A.~Gunawardana.
\newblock Evaluating recommendation systems.
\newblock In {\em Recommender systems handbook}, pages 257--297. Springer,
  2011.

\bibitem{teevan2004perfect}
J.~Teevan, C.~Alvarado, M.~S. Ackerman, and D.~R. Karger.
\newblock The perfect search engine is not enough: a study of orienteering
  behavior in directed search.
\newblock In {\em CHI'04}, pages 415--422, 2004.

\bibitem{Watts99}
D.~J. Watts.
\newblock Networks, dynamics, and the small-world phenomenon.
\newblock {\em American Journal of Sociology}, 105:493--527, 1999.

\bibitem{watts02}
D.~J. Watts, P.~S. Dodds, and M.~E.~J. Newman.
\newblock Identity and search in social networks.
\newblock {\em Science}, 296:1302--1305, 2002.

\bibitem{west2012human}
R.~West and J.~Leskovec.
\newblock Human wayfinding in information networks.
\newblock In {\em WWW'12}, pages 619--628, 2012.

\bibitem{white}
R.~W. White and S.~M. Drucker.
\newblock Investigating behavioral variability in web search.
\newblock In {\em {WWW '07}}, 2007.

\bibitem{bookcrossing}
C.-N. Ziegler, S.~M. McNee, J.~A. Konstan, and G.~Lausen.
\newblock Improving recommendation lists through topic diversification.
\newblock In {\em {WWW'05}}, 2005.

\end{thebibliography}
\end{document}